\documentclass[pre,floatfix,twocolumn]{revtex4}
\usepackage{graphicx} \usepackage{subfigure}
\usepackage{stdclsdv} \usepackage{array} \usepackage{amsmath}
\usepackage{amsfonts}

\newcommand{\av}[1]{\left\langle #1 \right\rangle}

\usepackage{color}

\begin{document}

\title{The free energy of a liquid when viewed as a population of overlapping clusters }

\author{Pierre Ronceray} \affiliation{LPTMS, CNRS, Univ. Paris-Sud,
  Universit\'e Paris-Saclay, 91405 Orsay, France \\}

\author{and Peter Harrowell} \affiliation{School of Chemistry,
University of Sydney, Sydney N.S.W. 2006, Australia}

\begin{abstract}
  The expression of the free energy of a liquid in terms of an
  explicit decomposition of the particle configurations into local
  coordination clusters is examined. We argue that the major
  contribution to the entropy associated with structural fluctuations
  arises from the local athermal constraints imposed by the overlap of
  adjacent coordination shells. In the context of the recently
  developed Favoured Local Structure model [Soft Matt. {\bf 11}, 3322
  (2015) ], we derive explicit expressions for the structural energy
  and entropy in the high temperature limit, compare this
  approximation with simulation data and consider the extension of
  this free energy to the case of spatial inhomogeneity in the distribution of local structures.
\end{abstract}

\maketitle

\section{ Introduction}

To distinguish crystals from liquids by the presence or absence of
periodic order, while operationally sound, is a lob-sided and
potentially misguided basis for the description of either
material. Clearly, the absence of periodicity in the liquid tells us
only what the liquid is not. As for the crystal, periodicity can
indeed describe the local arrangements of particles that determine the
energetics of the phase, but only for crystals with small unit
cells. A more generally useful distinction between the two phases is
that a crystal is structurally homogeneous while a liquid is
structurally heterogeneous.  In a survey of the 4264 structures of the
inorganic crystal structure data base, Daams and Villars~\cite{daams}
found that $88\%$ exhibited only 4 or less distinct local coordination
environments. A survey~\cite{steed} of molecular crystals found that
$92 \%$ consisted of only a single distinct coordination site. In
contrast, simulations of amorphous binary alloys regularly report
between 15 and 20 distinct local coordination polyhedra occurring with
significant frequency. This number can only increase in the 3 to 5
component alloys frequently used in bulk metallic glasses. An approach
to liquids based on the multiple possible local coordination
structures would, by default, also cover the crystal structures. This
consistent representation of the structure of both condensed phases
provides one the fundamental motivations for the subject of this
paper.

To refer to an arrangement of particles as a structure is to imply
that it can be intelligibly decomposed into some finite number of
components. A number of choices for these components have been
considered -- Voronoi polyhedral~\cite{bernal}, Delaunay
tessellations~\cite{ogawa}, common neighbours~\cite{jonsson}, bond
ring statistics~\cite{snook} and tetrahedrality~\cite{doye}. One of
the most common is the local coordination polyhedra~\cite{paddy} –
similar to a Delaunay vertex except that the neighbours are defined by
a cutoff separation rather than a geometric condition based on
space-filling. The uses of local coordination shells to construct
complex extended structures is the basis of the classic work of Frank
and Kasper~\cite{kasper}. The coordination polyhedra have the
appealing property that, for liquids comprised of particles that
interact only by short range potentials, the potential energy is
completely specified by the statistics of the coordination shells. The
other attractive feature of any structural description is data
compression. Indeed, the full description of the environment of a
particle involves $nd$ continuous degrees of freedom, where $n$ is the
number of neighbours of a site and $d$ the space dimension. In
contrast, once a dictionary of distinct structures has been
established, the local environment can be specified simply by
indicating in which of the $N_s$ possible environments the particle
is. Such structural characterizations, therefore, correspond to a
substantial coarse-graining (or ‘up scaling’~\cite{proc1}). What is
neglected in this description are the continuum of distortions – local
and global - that would effectively differentiate each local structure
and result in elastic-like coupling between local structures over an
extended range.  The discretization of the space of structures gained
through this coarse graining approximation is of considerable benefit
for the calculation of the liquid free energy – both in terms of the
mathematical practicalities (replacing integrals by sums) and in
providing some sort of intuition regarding the connection between the
particle properties and the stability of their liquid.

In a recent series of paper~\cite{proc1,proc2,proc3,proc4,proc5},
Procaccia and co-workers have developed an empirical approach to the
construction of a free energy in terms of an explicit decomposition of
liquid structures into local coordination clusters which they refer to
as quasi-species. They have applied this quasi-species construction to
binary atomic mixtures in 2D~\cite{proc2} and 3D~\cite{proc3}, a
molecular liquid~\cite{proc4} and a liquid characterised by
tetrahedral coordination~\cite{proc5}. As described in
ref.~\cite{proc1}, nearest neighbour clusters are categorized by the
total number of particles in the coordination shell. In a binary
mixture of large and small atoms, the mole fraction of clusters are
then denoted as $C_{s}(n)$ and $C_{l}(n)$ where $n$ is the number of
neighbours and $s$ or $l$ denote the identity of the central
particle. The composition of the coordination shell is not included in
this classification. Given the constraints of packing, there are only
a finite number of values of $n$ likely to be found about a particle of
a given species, so the number of types of cluster is finite ($\sim 5-6$ in
2D and 15-20 in 3D for the size ratios
studied~\cite{proc2,proc3}). Next, the average value of the structure
concentrations $\langle C_{\alpha}(n)\rangle (T)$ are calculated over
a range of temperatures and then inverted to give a cluster free
energy $F_{\alpha}(n,T)$ using

\begin{equation}
\langle C_{\alpha}\rangle (T)=\frac{\exp(-F_{\alpha}(m,T)/T)}{2\sum_{m}\exp(-F_{\alpha}(m,T)/T)}
\end{equation}

Note that this approach has factorized the free energy by neglecting
any contribution to the free energy from interactions between
clusters. The resulting free energies have been found to vary linearly
with T over the temperature range studied, a result that permits an
enthalpy and entropy to be defined for each cluster size, \emph{i.e.}

\begin{equation}
F_{\alpha}(n,T) \approx h_{\alpha}(n)-Ts_{\alpha}(n)
\end{equation}

The quasi-species construction is essentially descriptive; a strategy
for finding an optimal extrapolation of the observed $T$ dependence of
the cluster concentrations. A shortcoming of this
approach is the neglect of the correlations between
neighbouring local structures in the calculation of the
entropy. Adjacent coordination clusters
must overlap, so that a specific coordination on particle will impose
a constraint on the possible coordination shells of its
neighbour. Such constraints are geometric and hence
athermal and they result in complex correlations between the
  concentrations of various clusters. In this paper we present an
explicit treatment of this problem in the calculation of the
structural entropy. To this end, we employ a lattice model of liquids,
namely the Favoured Local Structure model, which we introduce in the
following Section. In Section 3 we derive an explicit expression for
the liquid free energy in the high temperature limit based on an
expansion in powers of the structure concentrations. A comparison
between the approximate theory and Monte Carlo calculations is
presented. In Section 4 we consider the extension of the treatment to
address the free energy of the solid-liquid interface.

\begin{figure*}
    \centering
    \includegraphics[width=1. \columnwidth]{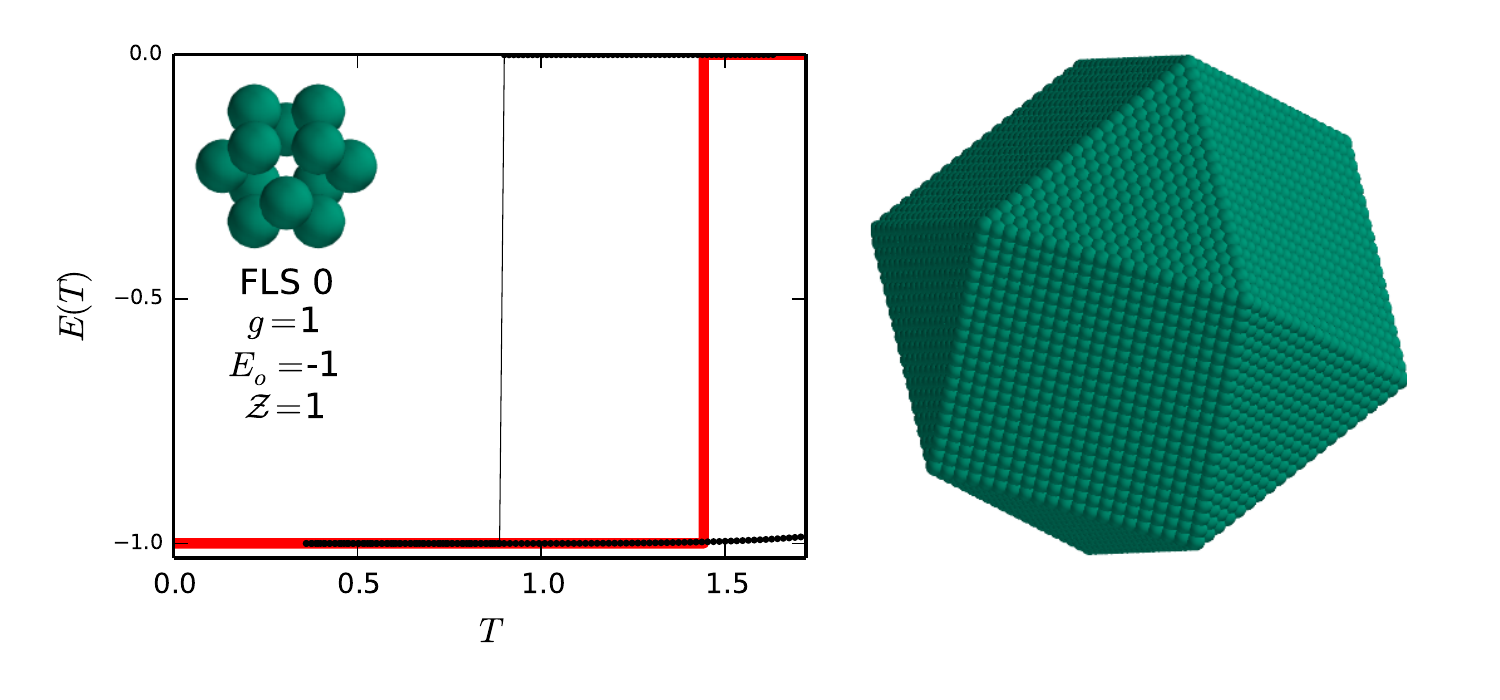}
    \includegraphics[width=1. \columnwidth]{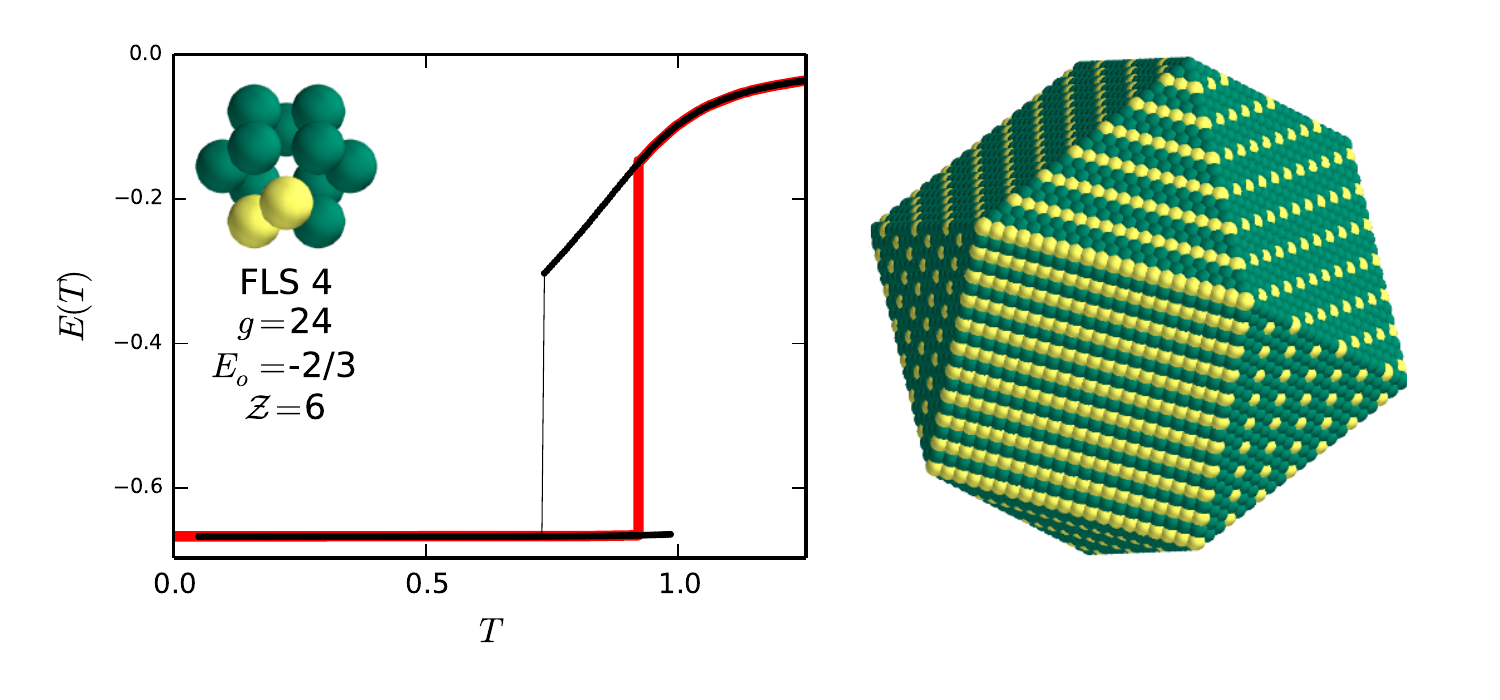}
    \includegraphics[width=1. \columnwidth]{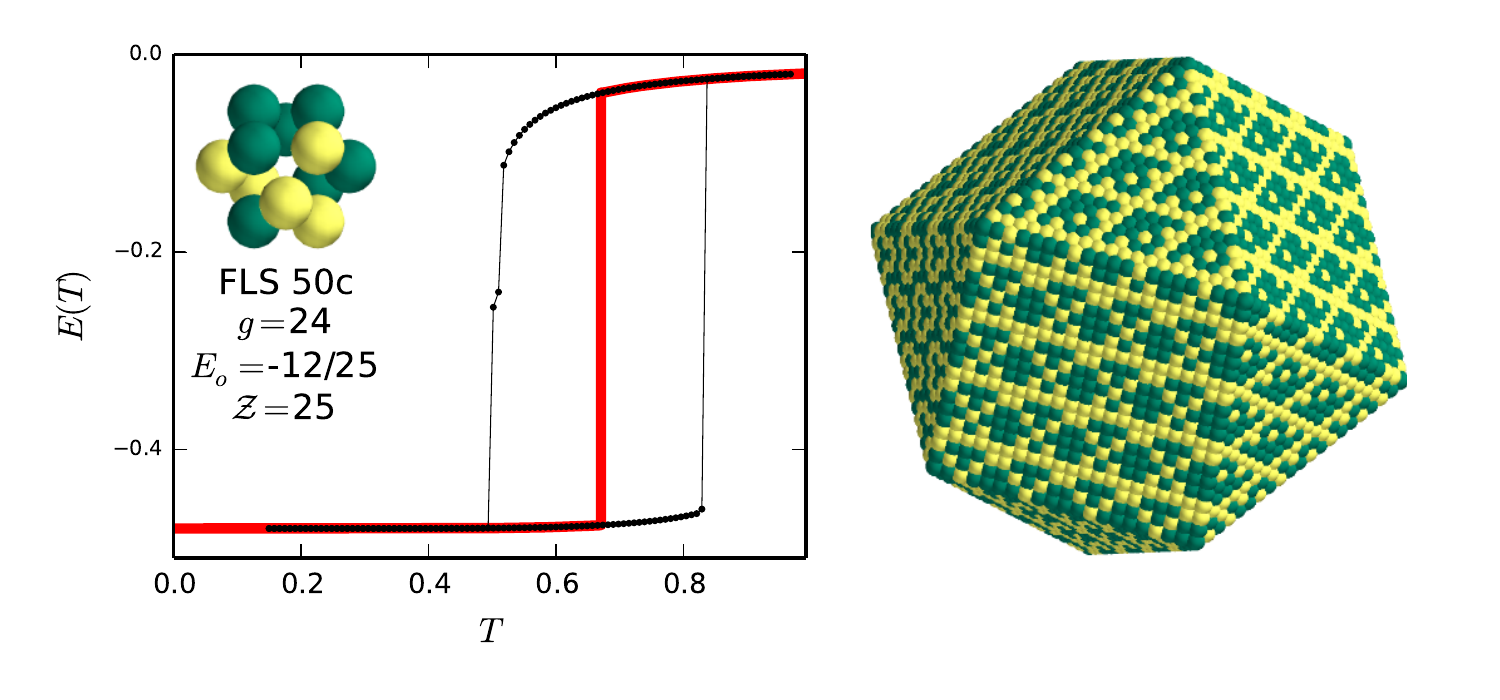}
    \includegraphics[width=1. \columnwidth]{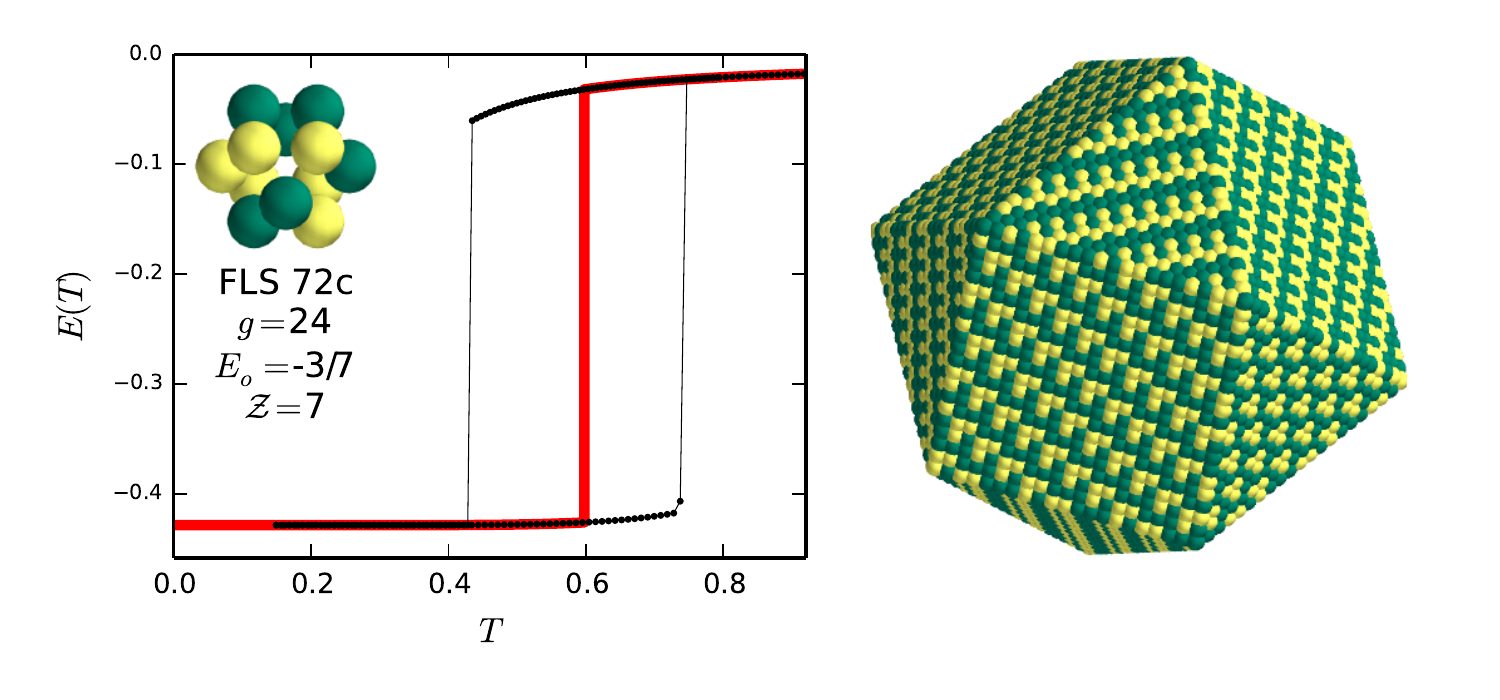}
    \caption{ Examples of different local structures in the 3D FLS
      model and the crystal groundstates resulting when they are
      selected as the single FLS. In each case we plot the associated
      energy vs temperature curves for heating and cooling Monte Carlo
      runs (black lines) along with the equilibrium curve (red
      line). The latter is obtained by comparing the free energies of
      both the liquid and crystal, obtained by thermodynamic
      integration from $T=\infty$ and $T=0$, respectively. The step
      transition in energy is indicative of a first order
      transition. Also provided are the rotational degeneracy $g$, the
      groundstate energy $E_o$ and the crystal unit cell size (as a
      number of lattice sites), $Z$.}
    \label{fig:FLS}
\end{figure*}

\section{The Favoured Local Structure Model}

The central challenge to developing a statistical mechanics based on
local structures is to handle the identification and counting of these
structures. Allowing particles to only occupy discrete positions in
space – \emph{i.e.} considering a lattice liquid – significantly
simplifies these difficulties. This was a major motivation behind our
development of the Favoured Local Structure (FLS) model. A lattice is
selected and every vertex assigned one of two values. We shall refer
to these two states as spin up or down.  (In a standard mapping, the
states can equivalently be thought of as atomic species A or B whose
identity can be interchanged in a Monte Carlo procedure.) The local
structure associated with a given site is determined by the
geometrical arrangement of the spins in the nearest neighbour
sites. We can enumerate the $N_s$ possible distinct local structure,
where two structures are considered distinct if they cannot be
transformed into each other by a rotation. For the 2D triangular
lattice, $N_{s}=14$ and for the 3D face centred cubic (fcc) lattice we
have $N_{s}=218$. (Note that in contrast with previous works~\cite{ronceray1,ronceray2,ronceray3,ronceray4,ronceray5}, we
distinguish here spin inverted structures.)  We assign an energy
$\epsilon_i$ to each local structure, typically selecting a small
number to have negative energies and hence be 'favoured' (see
Fig.~\ref{fig:FLS}). We shall refer to the model as the Favoured Local
Structure (FLS) model. The essence of our Hamiltonian is that the
energy attributed to each site is determined by the entire local
structure, not by the addition of pairwise interactions. A 2D version
of the FLS model has been studied~\cite{ronceray1} with the aim of
establishing the relation between the choice of FLS and the liquid
entropy~\cite{ronceray2} and the freezing
transition~\cite{ronceray3}. A novel liquid-liquid phase transition
was identified, in 2D, when the FSL was chiral~\cite{ronceray4}. The
extension of the FLS model to 3D based is the subject of
ref. ~\cite{ronceray5}. This paper includes a comprehensive survey of
the groundstates and phase transitions for all possible single FLS
choices on a fcc lattice. The numerical simulations of the model
reported below have been carried out using the Metropolis Monte Carlo
algorithm due to Bortz \emph{et al.}~\cite{bortz}, employing periodic
boundary conditions on an fcc lattice of typical size $24\times
24\times 24$.

\section{Structural Entropy: the High Temperature Expansion}

We now introduce a framework to study the implications of the local
structures on the liquid free energy.  Let $\vec{c}$ be the vector of
length $N_s$ whose components $\{c_{i}\}$ are the concentrations of
the $N_s$ distinct local structures.  The proportion of sites with
local environment $i$ is thus equal to $c_i$. As we include all
possible environments, it follows that $\sum_{i}c_{i} = 1$. At
infinite temperature, all spin configurations are equiprobable, and
thus the concentration vector $\vec{c}_{\infty}$ at infinite
temperature is given by
\begin{equation}
c_{\infty ,i}= g_{i}/2^{z}\label{eq:cinfty}
\end{equation}
where $z$ is the coordination number of the lattice ($z=12$ on the fcc
lattice). Here $g_{i}$ is the rotational degeneracy of $i$th
structure, which ranges between $1$ (for the all-\emph{up} and
all-\emph{down} structures) and $24$ (for any of the $136$ minimally
symmetric structures). For convenience we define the shifted
concentrations
\begin{equation}
\vec{\phi} = \vec{c} - \vec{c}_{\infty}
\label{eq:phi}
\end{equation}
\noindent such that $\langle \vec{\phi}\rangle =0$ as $T \rightarrow
\infty$.  We now consider a high temperature expansion of
the thermodynamic properties of the system.  Let $S(\phi)$ denote the
average entropy per site in a system restricted to this particular set
of structure concentrations. To second order in $\phi$ we can write

\begin{equation}
S(\vec{\phi})=S_{\infty}-\frac{1}{2}\vec{\phi}^{T} \cdot \hat{A} \cdot \vec{\phi} + O(\phi^3)
\label{eq:entropy}
\end{equation}
  							
\noindent such that the density of states in the structure space is
approximated by a multi-dimensional Gaussian distribution,

\begin{equation}
\Omega(\vec{\phi}) \approx \Omega _{\infty} \exp\left(-\frac{N}{2}\vec{\phi}^{T} \cdot \hat{A} \cdot \vec{\phi}\right)
\label{eq:omega}
\end{equation}
 							
\noindent where $N$ is the number of lattice sites in the system. Note
that $\Omega _{\infty} = 2$ in this model (corresponding to the two
spin states), and thus $S_{\infty}=\ln 2$.  Since in the FLS model (as
in many real world cases) the energy is completely determined by the
numbers of different local structures, the 'interactions' in this
model are entirely entropic and, in this high $T$ limit, encoded
within the {\it interaction matrix} $\hat{A}$. The {\it covariance
  matrix} $\hat{C}$ of this distribution is given by $\hat{C} =
\hat{A}^{-1}$.

Each instance of the model is characterised by specifying the energy
$\epsilon_{i}$ for each local structure $i$. We can write the energy
per site for a configuration described by a given structural
concentration $\vec{c}$ as

\begin{equation}
E(\vec{\phi})=\vec{\epsilon}\cdot\vec{c}=E_{\infty}+\vec{\epsilon}\cdot\vec{\phi}
\label{eq:energy}
\end{equation}
 								
\noindent where $E_{\infty}=\vec{\epsilon}\cdot\vec{c}_{\infty}$. At a
fixed temperature T, the free energy of the system with a given
structural concentration $\vec{\phi}$ is

\begin{eqnarray}
F(\vec{\phi},T) & =& E(\vec{\phi})-TS(\vec{\phi}) \\
\nonumber & = &E_{\infty} - TS_{\infty} +\vec{\epsilon}\cdot \vec{\phi}+\frac{T}{2}\vec{\phi}^{T} \cdot \hat{A} \cdot \vec{\phi}+O(\phi ^3)
\label{eq:Fe}
\end{eqnarray}

The average concentrations at temperature $T$ satisfy the condition
$\partial F/\partial \phi_{i} = 0$ resulting in the following
expression,
\begin{equation}
\vec{\phi}(T) = -\frac{1}{T}\hat{C}\cdot \vec{\epsilon} .
\label{eq:equil}
\end{equation}
This simple expression gives an approximation of the temperature
dependence of all local structures, which becomes exact in the
infinite temperature limit. The concentration of local structure $i$
thus depends directly on the energy of FLS $j$ through the term
$C_{ij} \epsilon_j /T$.  At this level of approximation, all
concentrations are proportional to $1/T$.  Substituting the
equilibrium structural concentrations, we have

\begin{eqnarray}
E(T) &= E_{\infty}-K/T  \label{eq:EK}\\
S(T) &= S_{\infty}-\frac{K}{2T^{2}} \label{eq:SK}\\
F(T)  &=E_{\infty}-TS_{\infty}-\frac{K}{2T} \label{eq:FK}
\end{eqnarray}

\noindent where
\begin{equation}
  K=\vec{\epsilon}^{T} \cdot \hat{C} \cdot
  \vec{\epsilon}\label{eq:K}
\end{equation}
is the key quantity that connects the energy of structures to the
thermodynamic properties of the liquid. Note that this quantity not
only involves the structures energies, but also the entropic
interaction between favoured structures.  Using Eq.~\ref{eq:EK} to replace $T$ by
$E$ we can write the microcanonical expression for the entropy,

 \begin{equation}
 S(E)=S_{\infty}-\frac{(E-E_{\infty})^{2}}{2K}
 \label{eq:micro}
 \end{equation}

 \noindent We note that a number of authors~\cite{entropy} have
 reported that the dependence of the configurational entropy on the
 inherent structure energy is well described by a quadratic function
 similar to that in Eq.~\ref{eq:micro}.

 To evaluate the quantity $K$ in the expressions above, we need the
 matrix elements of the covariance matrix $\hat{C}$. For a system of
 $N$ sites, the elements of $\hat{C}$ are related to the correlations
 between fluctuations of concentration of structures at infinite
 temperature through
\begin{equation}
C_{ij}=N\langle \phi_{i}\phi_{j}\rangle _{\infty}
\label{eq:C}
\end{equation}
Recalling that $\langle \phi_{i}\rangle _{\infty}=0$, a value of
$C_{ij} > 0$ corresponds to the case where concentrations of the two
local structures correlate positively, due to a geometrical affinity
for one another (\emph{i.e.} the presence of one structure facilitates
the formation of the other in its neighbourhood) while $C_{ij} < 0 $
indicates a geometric antipathy between structures $i$ and $j$. In
this sense, the matrix $\hat{C}$ (or its inverse $\hat{A}$)
encapsulates the essential constraints imposed by the local structures
in space, at least at the level of the pairwise overlap. Note that the
matrix $\hat{C}$ (unlike $K$) is independent of what energies are
assigned to the local structures and so a single matrix applies for
all possible versions of the FLS model.

In practice, we compute exactly the covariances in Eq.~\ref{eq:C} by
noting that it can be written as a finite sum of
probabilities. Indeed, working on the value of the
infinite-temperature correlation between concentrations
$\av{c_ic_j}_\infty$, we have:
\begin{eqnarray}
  \av{c_ic_j}_\infty & =  &\av{ \left( \frac{1}{N} \sum_{\vec{r_1}} \delta_{s(\vec{r_1}) ,i}\right) \!\! \left( \frac{1}{N} \sum_{\vec{r_2}}\delta_{s(\vec{r_2}) ,j}\right) }_\infty  \\
   & =  &\frac{1}{N^2}\sum_{\vec{r_1}}\sum_{\vec{r_2}}\av{\ \delta_{s(\vec{r_1}) ,i}\  \delta_{s(\vec{r_2}) ,j}\ }_\infty
\end{eqnarray}
where $\vec{r_1}$ and $\vec{r_2}$ are two lattice sites, and
$\delta_{s(\vec{r_1}) ,i}$ equals one if the site at position
$\vec{r_1}$ is in the local environment $i$, and zero otherwise.
Using translation invariance, we can set $\vec{r_1}=\vec{0}$ and
write:
\begin{eqnarray}
    \langle c_{i} c_{j}\rangle _{\infty} & =  & \frac{1}{N} \sum_{\vec{r}} \mathrm{Prob}\left( s_{\vec{0}} = i  \& s_{\vec{r}} = j \right)
\end{eqnarray}
where $\mathrm{Prob}\left( s_{\vec{0}} = i \& s_{\vec{r}} = j \right)$
denotes the joint probability that site $\vec{0}$ is in the local
structure $i$ and site $\vec{r}$ is in $j$. If these two structures do
not share any lattice site, they are independent at infinite
temperature, and the probability thus factorizes to the average
concentrations at infinite temperature $c_{\infty,i}
c_{\infty,j}$. Going back to the shifted concentrations $\phi_i = c_i
- c_{\infty,i}$, we can thus write:
\begin{equation}
   N \langle \phi_{i} \phi_{j}\rangle _{\infty}  =  \sum_{\vec{r} \sim \vec{0}} \mathrm{Prob}\left( s_{\vec{0}} = i  \& s_{\vec{r}} = j \right) - c_{\infty,i}c_{\infty,j}
\end{equation}
where $\vec{r} \sim \vec{0}$ means that we consider only positions
$\vec{r}$ whose local environment overlap with that of $\vec{0}$.  We
can rewrite this expression
\begin{equation}
   C_{ij}  =  c_{\infty,i} \sum_{\vec{r} \sim \vec{0}} \left( \mathrm{Prob}\left(s_{\vec{r}} = j | s_{\vec{0}} = i \right) - c_{\infty,j} \right)
\end{equation}
where $\mathrm{Prob}\left(s_{\vec{r}} = j | s_{\vec{0}} = i \right)$
denotes the conditional probability to find site $\vec{r}$ in local
structure $j$, given that $\vec{0}$ is in $i$. This last expression is
tractable, and can be computed by exact counting of all possible ways
to successfully fit local structure $j$ at all positions overlapping
with the origin. In practice, this required the evaluation of
approximately $5\times10^7$ overlaps to compute the full covariance
matrix in the 3D model.

\begin{figure}[th]
    \centering
    \includegraphics[width=\columnwidth]{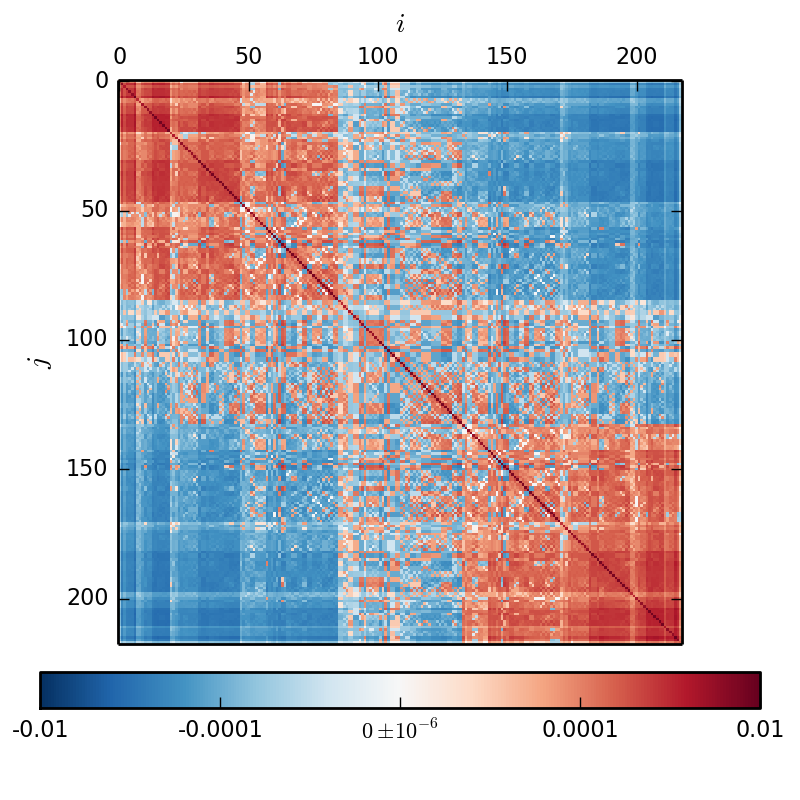}
    \caption{ A colour coded depiction of the values of the elements
      of the covariant matrix $\hat{C}$. Negative and positive elements are in
      shades of blue and red, respectively. These shades are
      log-scaled with a cut region around zero, as indicated by the
      colorbar. The local structures, listed by a numerical index
      running from 1 to 218, are sorted by their number of up spins
      (hence the block structure of this matrix, as local structures
      with similar spin composition will tend to attract each
      other). Within each block, they are sorted with increasing
      geometrical multiplicity $g_i$.}
    \label{fig:covar-big}
\end{figure}

The matrix $\hat{C}$ is singular, as the unphysical eigenvector
$\vec{\phi_\mathrm{unphys}}=(1,1,\dots 1)$, associated to
concentrations that do not sum to unity, corresponds to a zero
eigenvalue of $\hat{C}$. Surprisingly, we found that in the 3D FLS
model, two other eigenvalues vanish. These non-trivial zero modes
correspond to forbidden directions $\vec{\phi_\mathrm{forbidden}}$ in
the concentration space: these compositions indeed incur an infinite
entropy cost. Physically, they correspond to local structures whose
neighbourhood \emph{must} include some other type of structure: it is
impossible to simultaneously increase their concentration, and
decrease the concentration of all their ``followers''. We disregard
these oddities in the following. To obtain the interaction matrix
$\hat{A}$, we eliminate the zero modes by singular value
decomposition, using the Moore-Penrose pseudo-inverse of $\hat{C}$.

In Fig.~\ref{fig:covar-big} we present a colour-coded representation
of the covariance matrix. The local structures are indexed in blocks
characterized by their number of up spins. So the all down local
structure is given an index $0$, the local structure with a single up
spin is structure $1$, the four structures with $2$ up spins are
structures $2$ to $5$, arranged in order of increasing rotational
multiplicity $g$, and so on. The gross features of $\hat{C}$, evident
in Fig. ~\ref{fig:covar-big}, result from this choice of organizing
the local structures. The blocks of positive covariance seen in the
top left and bottom right corners correspond to the positive overlap
of local structures that both consist of predominantly down or up
spins, respectively. Similarly, the blocks of negative covariance in
the top right and bottom left corners are the negative overlaps that
result between a pair of structures whose majority spins are opposite
in sign. The boundaries of these corner blocks correspond to the last
of the 5 up (or down) spin structures. The local structures with equal
numbers of up and down spins make up the central cross-like
feature with its complex patchwork of positive and negative
covariance. The diagonal consists of the variances
$N\langle\phi^{2}_{i}\rangle$ which must be positive.

\begin{figure}[th]
    \centering
    \includegraphics[width=\columnwidth]{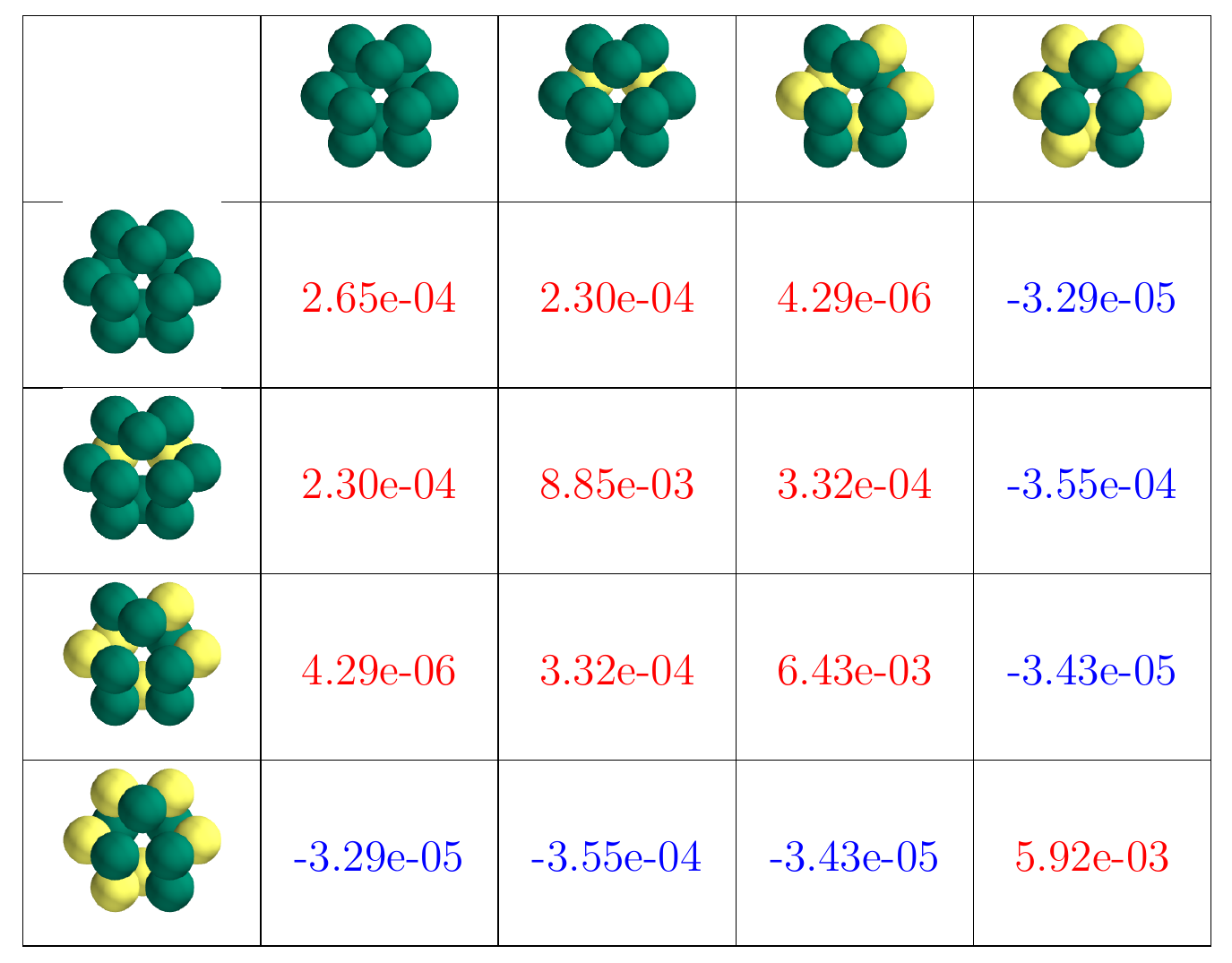}
    \caption{ A selection of the covariances represented in
      Fig.~\ref{fig:covar-big}. The numerical values of the
      covariances are provided in the matrix.}
    \label{fig:covar-small}
\end{figure}

To appreciate the detail of the covariance matrix we have extracted
the covariances of just four local structures and presented them in
Fig.~\ref{fig:covar-small}. Clearly evident is the tendency for
structures to have positive covariance when their majority spin is the
same.

\begin{figure}[th]
    \centering
    \includegraphics[width=\columnwidth]{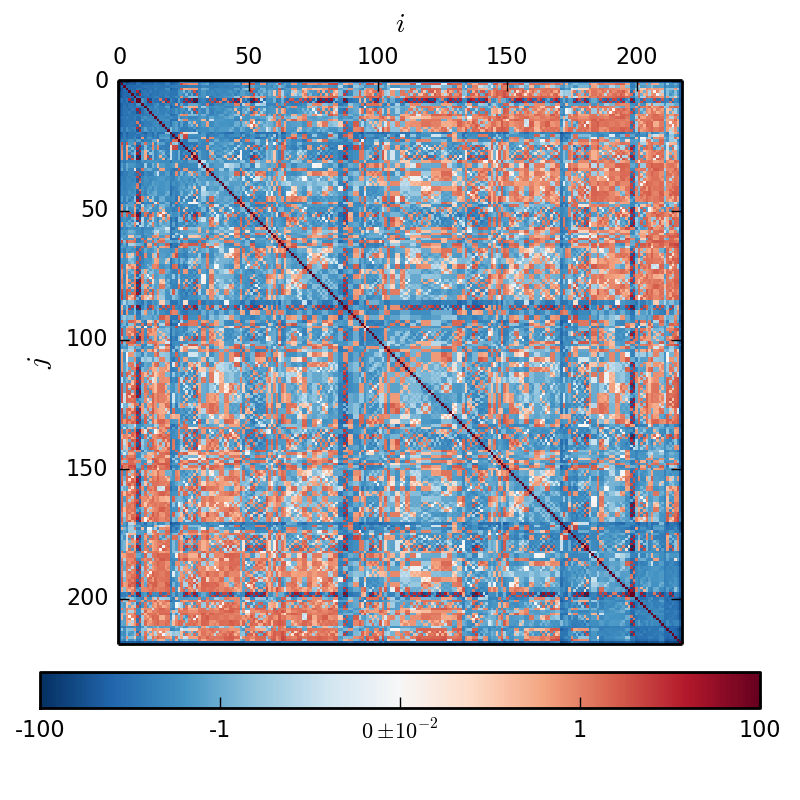}
    \caption{  A colour coded depiction of the values of the elements
      of the interaction matrix $\hat{A}$. Negative and positive elements are in
      shades of blue and red, respectively.  }
    \label{fig:A}
\end{figure}

The block-like structure evident in $\hat{C}$ is largely absent in
$\hat{A}$ as shown in Fig.~\ref{fig:A}. The interaction matrix
exhibits a striking diagonal organization. Matrix elements close to
the diagonal, i.e. corresponding to pairs of structures on slightly
different from one another, we generally find negative values of
$A_{ij}$ which imply structural compatibility. As we move further from
the diagonal, the increasing difference in the sign of spins of the
two local structures results in a positive (\emph{i.e.} antagonistic)
interaction. The diagonal values stand apart as they must be strongly
positive to ensure that the high $T$ limit corresponds to a maximum in
entropy.

\begin{figure}[th]
    \centering
    \includegraphics[width=\columnwidth]{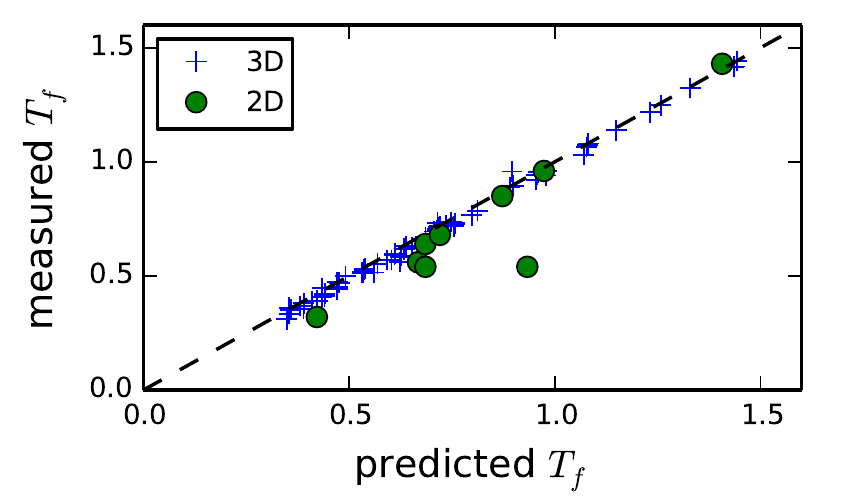}
    \caption{ A scatter plot of simulated melting points vs
      predictions from Eq.~\ref{eq:Tm} for the single FLS cases in 3D
      and 2D. The dashed line corresponds to the case where the
      predicted and the actual values are equal.}
\label{fig:Tm}
\end{figure}

How well does the high $T$ approximation fare in estimating properties
of the FLS model? The freezing point $T_f$ is obtained by requiring
that the free energy of the liquid, Eq.~\ref{eq:FK}, is equal to the
free energy of the crystal. The latter quantity can be reasonably
approximated by zero entropy and a groundstate energy $E_o$. Doing so
we arrive at the following expression for $T_f$,

\begin{equation}
T_{f} = \left({\frac{E_{\infty}-E_o}{2S_{\infty}}}\right) \left(1+\sqrt{1-\frac{2KS_{\infty}}{(E_{\infty}-E_o)^{2}}}\right)
\label{eq:Tm}
\end{equation}
With the calculated covariances and the crystal groundstate energies
we can use Eq.~\ref{eq:Tm} to predict the melting points for various
choices for favoured local structures and compare these predictions in
Fig.~\ref{fig:Tm} with those values obtained from the MC
simulations. The approximate expression for $T_f$ provides an
excellent prediction of the actual freezing point of a single FLS in
the 3D and 2D models (with one exception in the latter case). This
success is a reflection of the fact that liquids generally do not
accumulate much local crystalline order before freezing, a direct
consequence of the entropy cost of a single FLS. Small FLS
concentrations are the basis for the expansion used to derive
Eq.~\ref{eq:Tm}.

\begin{figure*}
    \centering
    \includegraphics[width=0.9 \textwidth]{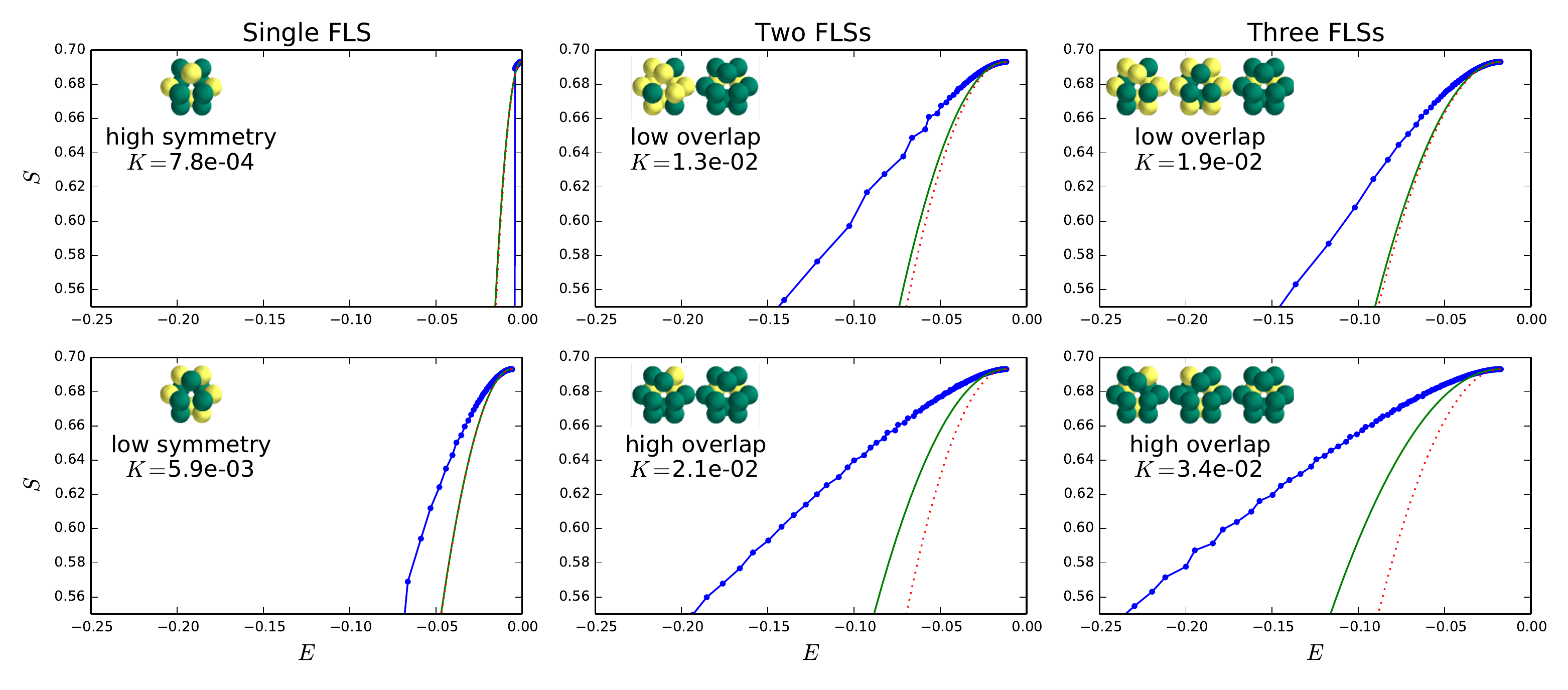}
    \caption{ The liquid entropy as a function of energy calculated by thermodynamic integration from simulations (blue circles) and using the approximate expression in Eq.~\ref{eq:micro} using the exact expression for $K$ (Eq.~\ref{eq:K}) (green line) and the approximate $K$ (Eq.~\ref{eq:Kapprox}) (red dots). Examples with one, two and three FLS's are presented along with the value of $K$ from Eq.~\ref{eq:K}.}
\label{fig:entropy}
\end{figure*}

More informative is the entropy as a function of energy. In
Fig.~\ref{fig:entropy} we compare the actual entropy as a function of
energy, calculated using thermodynamic integration~\cite{ronceray2},
with the approximate quadratic form in Eq.~\ref{eq:micro}, for a
variety of cases involving one, two or three FLS's. The key
contributions of liquid structure to the stability (\emph{i.e.} high
entropy) of liquids are evident. Low symmetry FLS's and high overlap
(of an FLS with itself and, in the case of multiple favoured
structures, between FLS's) both contribute to larger $K$'s and, hence,
to higher liquid entropies for given energy. The high $T$
approximation performs reasonably for the case of a single FLS,
consistent with the similar success for the approximate expression for
$T_f$ discussed previously. We find that Eq.~\ref{eq:micro}, while
continuing to work at high energies, consistently underestimates the
liquid entropy. This means that the pairwise interactions alone
generally underestimate the capacity of structures to ``pack'' in
space. Why this should be so is not obvious and represents an
interesting question for future study. 

Some attempts at simplifying the expression for the entropy are
informative. Consider the following naive treatment of the problem in
which we let each site be assigned one of the possible local
structures independently of their neighbours. The result is a massive
overestimation of the entropy as we have replaced the true number of
possible states for a site, \emph{i.e.} $2$, with the very much larger
number $2^z$ of possible local structures (with their rotational
variants). The lesson here is that local structures inherently span
over several degrees of freedom, and so they cannot decouple
completely. We can however obtain a useful approximation by assuming
that overlaps between structures are ``neutral'', \emph{i.e.} that the
presence of an FLS at a given site does not affect the probability of
having an FLS at nearby sites. In this case, the entropy depends only
on the symmetry properties of the local structures, which are encoded
in the infinite-temperature concentration $c_{\infty,i}$ of FLS $i$
(Eq.~\ref{eq:cinfty}). It is thus much simpler to compute: indeed, in
this case, we neglect off-diagonal terms in the covariance matrix, and
$K$ can be simply approximated as
\begin{equation}
K \approx \sum_{i}^{N_s}\epsilon_{i}^{2}c_{\infty,i}
\label{eq:Kapprox}
\end{equation}
The resulting approximation to $S(E)$, obtained by substituting this
approximate $K$ into Eq.~\ref{eq:micro}, has been plotted in
Fig.~\ref{fig:entropy}. Eq.~\ref{eq:Kapprox} provides an accurate
estimation of $K$ when overlap between the FLS is small but clearly
underestimates $K$ (\emph{i.e.} underestimates the second order
expression for $S(E)$) when the overlap is significant. 

Finally, we note that $S(E)$ is a smoothly varying function of $E$
and, as such, can be reasonably described by Eq.~\ref{eq:micro} across
much of the energy range {\it if} we allow $K$ to be an effective,
adjustable parameter $K_{\mathrm{eff}}$ fitted through the whole
energy range. Capturing, as it does, the compounded complexities of
local structure energies, the nonlinear contributions of overlap to
the entropy and the influence of multiple FLS's, $K_{\mathrm{eff}}$
should provide a useful quantity in the comparison of the role of
local structure in different liquids, with the drawback that it is no
longer a coefficient in an exact expansion, and thus cannot be
computed exactly from simple rules.
 
\section{The Susceptibility of a Liquid to Fluctuations in Local Structure}

Structure in liquids matters most when we are interested in how that
liquid will respond to some perturbation. A crystal surface, for
example, will grow if the perturbations in local crystal order that it
induces in the adjacent liquid are unstable and result in the
propagation of the interface. Crystal nucleation theory seeks to
describe the unstable wing of the (metastable) equilibrium
distribution of the local structures in the supercooled
liquid. Structural fluctuations in liquids involving more than
pairwise correlations have become directly observable with techniques
such as fluctuation electron microscopy~\cite{voyles}. To consider
fluctuations in structure we need to generalize our treatment of the
free energy to include spatial heterogeneities in structure,
\emph{i.e.} $\vec{\phi}(\vec{r})$.  The energy expression is easily
generalized,

\begin{equation}
E[\vec{\phi}]=E_{\infty} + \frac{1}{N}\sum_{\vec{r}}\vec{\phi}(\vec{r})\cdot \vec{\epsilon}
\end{equation}

\noindent where $N$ is the number of sites. For the entropy, we shall again resort to our high $T$ (and, hence, small $\phi$) expansion. Now $S[\vec{\phi}]$   is a functional of $\vec{\phi}(\vec{r})$  and so the Taylor expansion is of the form

\begin{equation}
S[\vec{\phi}]\approx S_{\infty}-\frac{1}{2N^{2}} \sum_{\vec{r}_{1}} \sum_{\vec{r}_{2}} \vec{\phi}^{T}(\vec{r}_{1})\cdot\hat{A}(\vec{r}_{1}-\vec{r}_{2})\cdot\vec{\phi}(\vec{r}_{2})
\end{equation}
 				
\noindent and the equilibrium condition is

\begin{equation}
N\frac{\delta F}{\delta \vec{\phi}(\vec{r}_{1})}=\vec{\epsilon}+\frac{T}{N}\sum_{\vec{r}_{2}}\hat{A}(\vec{r}_{12})\cdot\vec{\phi}(\vec{r}_{2})=0
\end{equation}
 					
\begin{figure}
    \centering
    \includegraphics[width= \columnwidth]{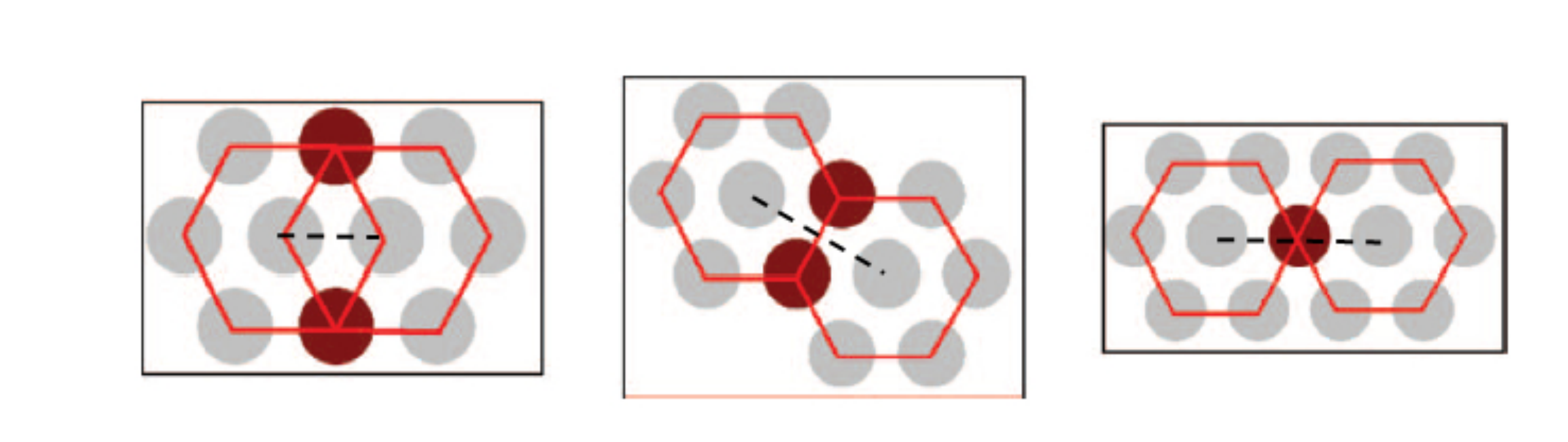}
    \caption{ The extent of overlap between a pair of local structures
      in the 2D triangular lattice. The dashed line indicates the
      three separations between sites for which the structure overlap
      is non-zero.}
\label{fig:overlap}
\end{figure}

The spatial dependence of the interaction matrix $\hat{A}$ elements
provides the contains the essential information about the locality of
the overlap of local coordination shells. The spatially resolved covariance matrix $\hat{C}(\vec{r})$ is given by

\begin{equation}
C_{ij}(\vec{r})=N\langle \phi_{i}(0)\phi_{j}(\vec{r}) \rangle
\label{eq:Cr}
\end{equation}
and corresponds to the two-point correlation between local structures. In the high $T$ limit, these structural correlation functions can be calculated by an extension of the approach used to produce Fig.~\ref{fig:covar-big}. This analysis is left for future work. The connection between the covariance matrix $\hat{C}(\vec{r}_{1},\vec{r}_{2})$ and the entropy via the interaction matrix $\hat{A}(\vec{r}_{1},\vec{r}_{2})$ is not as straightforward as it is for the homogeneous case since the relation between the two matrices will involve some sort of nonlocal convolution. Allowing for this complexity, we can make the following observation. The spatial correlations of the covariance matrix $\hat{C}(\vec{r})$ are restricted to the short distance set by the extent of overlap of local structures. In Fig.~\ref{fig:overlap} we show how the overlap of local coordination shells in the 2D triangular lattice only extends out to second nearest neighbour sites. A similar length is obtained in the fcc lattice in 3D. For separations beyond this overlap limit, $C_{ij}(\vec{r})=0$. While we expect that the interaction matrix $\hat{A}(\vec{r})$ will exhibit a similar spatial extent of its influence, the exact relation must wait establishing a clearer picture of the relation between the entropic interactions between local structures and their spatial correlations.

\section{Discussion}
We have presented a formalism in which the free energy of a liquid is
expressed explicitly in terms of the local coordination structure. The
novel feature of this theory is the treatment of the structural
contribution to the entropy which allows us to describe the entropic
consequences of the geometry of an individual favoured local structure
and the impact of multiple favoured local structures. The connection
between structure and entropy arises from the overlap of the local
coordination shells and the degree of local structural constraint this
imposes. While we have restricted ourselves to the high $T$
approximation, the overlap terms are athermal and so much of the
details of the affinities between pairs of local structures is
retained in the interaction matrix $\hat{A}$ and its inverse, the
covariance matrix $\hat{C}$. We considered only the expansion of the
entropy to quadratic order in the energy. In the context of single FLS
liquids, we previously described how higher order terms in this
expansion can be computed as sums of overlaps of larger
clusters~\cite{ronceray3}. However, we found that the complexity of
these terms increases very rapidly, for only moderate improvement of
the entropy estimate (data not shown).

The extension of the formalism to spatial inhomogeneities in the field
of structural concentrations has been introduced. This spatial
variation, if coupled to a free energy with terms of order greater
than the quadratic ones that characterise our high $T$ approximation,
will provide the basis for a theory of the crystal-liquid interfacial
free energy and the free energy of crystal nuclei in which, for the
first time, the influence of non-crystalline local structures can be
explicitly mapped out.

{\bf Acknowledgements} PH acknowledges the financial support of the
Australian Research Council. PR is supported by ``Initiative Doctorale
Interdisciplinaire 2013'' from IDEX Paris-Saclay.  Figures realized
with Matplotlib~\cite{matplotlib} and
Mayavi2~\cite{mayavi}.

\end{document}